\documentclass[aps,prxquantum,reprint,longbibliography]{revtex4-2}

\usepackage{amsmath,amssymb,mathtools,bm,amsthm}
\usepackage{graphicx}
\usepackage{booktabs}
\usepackage{braket}
\usepackage{hyperref}
\usepackage{xcolor}
\usepackage{enumitem}

\usepackage{algorithm}
\usepackage{algorithmicx}
\usepackage{algpseudocode}

\newtheorem{theorem}{Theorem}

\newtheorem{corollary}{Corollary}[section]

\begin{document}

\title{Residual-Algebraic Design of Symmetry-Preserving Variational Quantum Circuits}

\author{Sahinur Reja}
\affiliation{Department of Physics, Jadavpur University, Kolkata 700032, West Bengal India}

\date{\today}

\begin{abstract}
Symmetry-preserving (SP) variational quantum circuits restrict evolution to physically relevant sectors of Hilbert space, but symmetry preservation alone does not guarantee accessibility within those sectors. We show that a restricted generator set can possess a residual dynamical Lie algebra whose commutant further decomposes the target symmetry sector into invariant blocks. The associated block populations are conserved and impose a kinematic upper bound on the fidelity with a target state. We then invert this diagnostic into a circuit-design principle: candidate SP generators can be screened before variational optimization by asking whether they break the residual commutant, which invariant blocks they couple, and what adjoint module they generate under the residual algebra. For two orthogonal residual blocks we prove that a single elementary real inter-block rotation generates the full orthogonal algebra. We further derive a complex escape criterion: if the real and imaginary parts of an inter-block seed are independent over the residual intertwiner algebra, the generated module is the full complex coupling space and the enlarged algebra becomes special unitary. A connected residual-block graph yields a natural multi-block generalization and suggests minimum-cost spanning-tree circuit design. We demonstrate this design principle in the XXZ and Hubbard models, where residual-algebra screening identifies SP circuit extensions that remove the relevant accessibility obstructions. These examples establish a unified route from residual-symmetry diagnosis and fidelity bounds to predictive quantum circuit design.
\end{abstract}

\maketitle

\section{Introduction}
\label{sec:introduction}

Variational quantum algorithms (VQAs) provide a natural hybrid
quantum--classical framework for addressing quantum many-body problems
on present and near-term quantum processors
\cite{Peruzzo2014,Cerezo2021,Tilly2022}.  Among them, the variational
quantum eigensolver (VQE) has emerged as a leading approach for
approximating ground-state properties of molecular, spin, and fermionic
Hamiltonians.  Instead of representing an exponentially large many-body
wave function explicitly, VQE prepares a parametrized quantum state
$\ket{\psi(\boldsymbol{\theta})}$ and optimizes its parameters by
minimizing the variational energy
\begin{equation}
   E(\boldsymbol{\theta})
   =
   \bra{\psi(\boldsymbol{\theta})}
   H
   \ket{\psi(\boldsymbol{\theta})}.
\end{equation}
The practical performance of the method therefore depends critically on
the variational ansatz.  A useful circuit must be sufficiently expressive
to contain the physically relevant states, while remaining shallow,
trainable, and compatible with the symmetries and connectivity of the
underlying problem.

This balance between expressibility and trainability has motivated a
wide range of ansatz-design strategies.  Generic highly expressive
circuits can suffer from increasingly flat optimization landscapes and
barren plateaus \cite{McClean2018,Cerezo2021Barren,Ragone2024},
whereas strongly constrained circuits may remain trainable but fail to
contain the desired state.  Problem-inspired and Hamiltonian
variational ans\"atze provide an appealing intermediate approach by
constructing the circuit from operators related to the target
Hamiltonian \cite{Wecker2015,Wiersema2020,Jattana2022,Bosse2022}.
Such constructions have been explored for interacting spin models,
including Heisenberg and XXZ systems
\cite{CerveraLierta2021,Jattana2022,Bosse2022}, as well as for
Fermi--Hubbard systems
\cite{Seki2022,Mineh2022,Martin2022}.  Adaptive methods such as
ADAPT-VQE instead grow the circuit iteratively from an operator pool,
providing another route to compact problem-tailored ans\"atze
\cite{Grimsley2019,Tang2021,Grimsley2023}.

An important complementary principle is to encode known physical
symmetries directly into the circuit.  Particle number, total
magnetization, spin, or other conserved quantities can be enforced at
the level of the generators so that the variational evolution remains
inside the physically relevant sector
\cite{Gard2020,Barron2021,Seki2022}.  Restricting the search to such a
sector can substantially reduce the effective Hilbert space and suppress
unphysical directions.  Symmetry adaptation, however, introduces a
subtle design problem: preserving the required physical symmetry does
not imply that the circuit explores the entire corresponding symmetry
sector.  A particular generator set can preserve additional quantities
that are not symmetries of the target problem.  The resulting
\emph{residual ansatz symmetries} further partition the intended
symmetry sector into invariant components and can prevent the circuit
from reaching the target state even in the limit of arbitrarily large
depth.

This distinction is closely related to the algebraic structure of
parametrized quantum circuits.  Given anti-Hermitian circuit generators
their dynamical Lie algebra (DLA)
determines the connected dynamical group generated by the circuit and
therefore constrains its reachable orbit
\cite{Schirmer2001,Altafini2002,DAlessandro2007,Larocca2022}.
Lie-algebraic methods have consequently become increasingly important
for understanding expressibility and trainability in variational
circuits.  In particular, the dimension and representation structure of
the DLA have been related to gradient concentration and barren plateaus
\cite{Larocca2022,Ragone2024,Fontana2024}, while systematic
classifications have shown that the generated algebra can depend
sensitively on the locality and interaction topology of the available
generators \cite{Wiersema2024,KokcuGraph2026}.  Thus, two circuits that
preserve the same physical symmetries and contain comparable numbers of
parameters can nevertheless possess qualitatively different reachable
many-body manifolds.

In our recent work, we showed that even modest changes in the connectivity
of symmetry-preserving excitation circuits can produce large changes in
variational accessibility and VQE performance in Heisenberg- and
Hubbard-type systems \cite{Reja2026Connectivity}.  These results indicate
that circuit connectivity is not merely a hardware constraint, but can
control the accessible many-body manifold.  The present work develops the
algebraic mechanism underlying this connectivity dependence and turns it
into a predictive circuit-design principle.

Algebraic restrictions also arise in adaptive circuit construction, where the completeness and symmetry structure of the operator pool determine the reachable state manifold. In ADAPT-type approaches, incomplete or symmetry-incompatible pools can lead to convergence roadblocks even when the individual operators are physically motivated \cite{Shkolnikov2023}. Moreover, adaptive construction can become costly because operator selection requires repeated gradient evaluation over the pool together with reoptimization of an increasingly large circuit. Related work has therefore emphasized minimal or algebraically complete generator pools for controllability and efficient state preparation \cite{Viswanathan2025}, while recent geometric analyses have formulated reachability directly in terms of the dimension and orbit structure of the dynamical group \cite{Ngairangbam2026}. These developments establish that circuit architecture is fundamentally a choice of generator algebra and its associated orbit.

These studies largely address the forward question: what algebra and reachable manifold follow from a chosen generator set? For circuit design, however, the inverse problem is more consequential. Once a SP ansatz is known to be restricted, one would like to identify the residual invariant responsible for the obstruction, quantify how strongly it limits the target-state fidelity, and determine which physically admissible generator removes it most efficiently. Full Lie-algebra construction can itself become expensive at large Hilbert-space dimension, motivating scalable diagnostics based on commutants, residual block structure, and adjoint-module growth. Such diagnostics can screen candidate circuit modifications before repeated variational optimization. Increasing depth alone cannot remove an algebraic obstruction, since it explores the same orbit rather than changing it; a predictive design framework should therefore identify useful architecture before costly VQE optimization is undertaken.

Here we develop a residual-algebra framework that turns variational accessibility into a circuit-design problem. For a given SP generator set, the commutant reveals residual invariant sectors and conserved block populations that restrict the accessible variational orbit and impose depth-independent fidelity bounds. We then use this structure constructively to classify candidate generators by the residual sectors they connect and by the algebraic growth they induce. This leads to a predictive strategy for selecting circuit modifications before costly VQE optimization. Within this framework, we establish two algebraic completion results that characterize when an added inter-sector coupling is sufficient to merge residual blocks and enlarge the accessible algebra.

We demonstrate the resulting design principle in interacting spin and fermionic models, where restricted generator topologies produce nontrivial accessibility obstructions that are removed by suitably chosen additional connections or correlated generators. The analysis shows that circuit improvement is governed not simply by depth or parameter count, but by whether the added generator destroys the relevant residual invariant and induces the required algebraic completion. Residual algebra therefore provides both a diagnosis of variational failure and a constructive route to circuit architecture: identify the residual obstruction, determine the coupling needed to remove it, and build the minimal SP generator set capable of accessing the target manifold.
\section{Residual Algebra, Invariant Blocks, and Fidelity Constraints}
\label{sec:residual_framework}
\begin{figure}[t]
   \centering
   \includegraphics[width=\columnwidth]{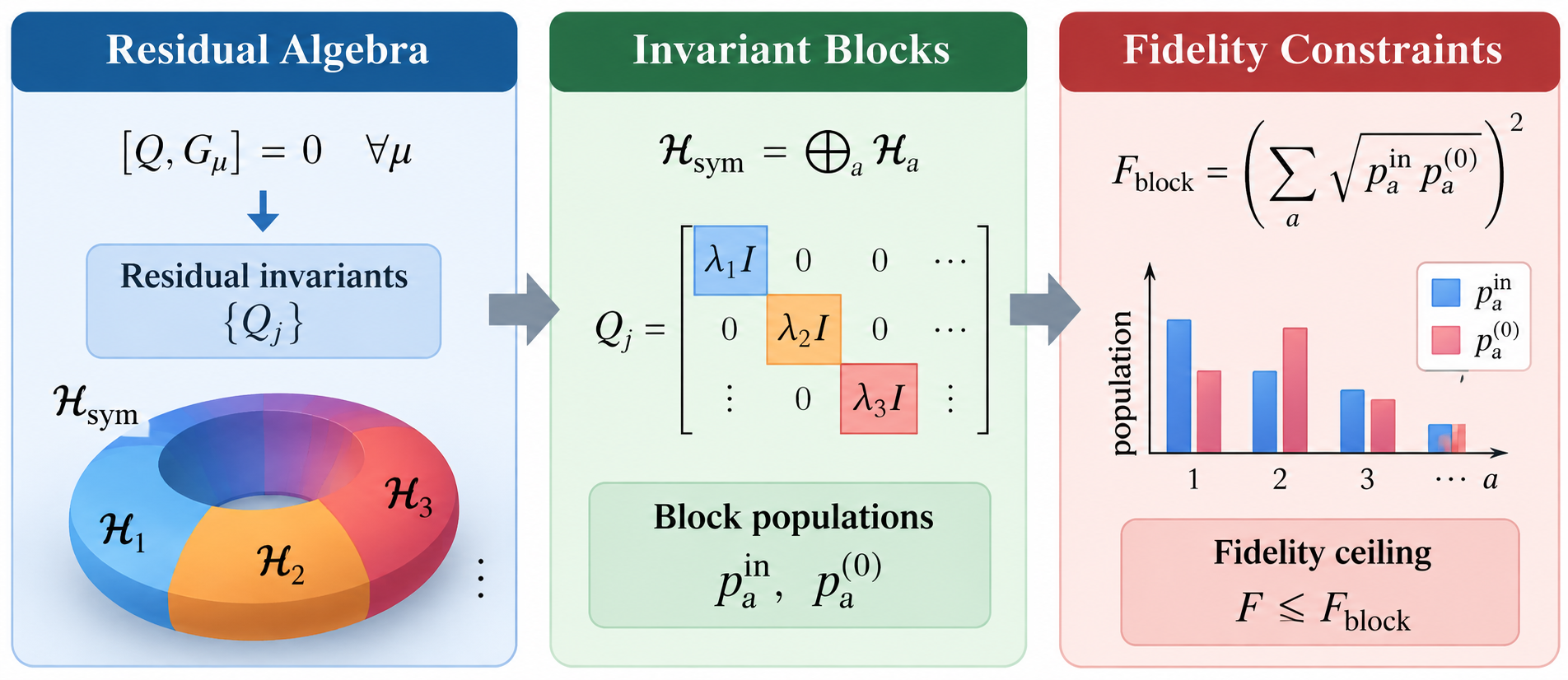}
   \caption{Residual algebra partitions the physical symmetry sector into invariant blocks whose conserved populations restrict the accessible orbit and impose a fidelity ceiling.}
   \label{fig:concept}
\end{figure}

\subsection{Variational orbit and residual algebra}
\label{subsec:orbit}

Consider a SP parametrized circuit constructed from a
fixed set of anti-Hermitian generators
$\mathcal G=\{G_\mu\}$,
\[
   U_L(\boldsymbol{\theta})
   =
   \prod_{\ell=1}^{L}
   \prod_\mu
   e^{\theta_{\ell\mu}G_\mu},
\]
where $L$ denotes the circuit depth.  The generator set determines the
dynamical Lie algebra (DLA)
\begin{equation}
   \mathfrak g
   =
   \operatorname{Lie}\{G_\mu\},
   \label{eq:dla_definition}
\end{equation}
defined as the smallest real Lie algebra containing the original
generators and all nested commutators generated from them.  For an
initial state $\ket{\psi_{\rm in}}$, the corresponding Lie group
$e^{\mathfrak g}$ defines the algebraically accessible orbit
$\mathcal O_{\psi_{\rm in}}
=\{U\ket{\psi_{\rm in}}:U\in e^{\mathfrak g}\}$.
The finite-depth variational manifold $\mathcal M_L$ explored by a
particular circuit parameterization is generally only a subset of this
orbit,
\begin{equation}
   \mathcal M_L
   \subseteq
   \mathcal O_{\psi_{\rm in}}
   \subseteq
   \mathcal H_{\rm sym},
   \label{eq:manifold_orbit_sector}
\end{equation}
where $\mathcal H_{\rm sym}$ is the Hilbert-space sector selected by
the intended physical symmetries.  Increasing depth can enlarge
$\mathcal M_L$ within $\mathcal O_{\psi_{\rm in}}$, but it cannot
enlarge the orbit itself as long as the generator set remains
unchanged.  Thus, a circuit can acquire more parameters and greater
finite-depth flexibility while remaining algebraically confined to a
restricted subset of the physical symmetry sector.

\subsection{Residual commutant, invariant blocks, and conserved populations}
\label{subsec:commutant_blocks}

Restrictions of the accessible orbit can be diagnosed through the
commutant of the generated algebra,
$\operatorname{Comm}(\mathfrak g)
=\{Q:[Q,G]=0,\ \forall G\in\mathfrak g\}$.
A nontrivial commutant operator represents an invariant of the ansatz
dynamics beyond the physical symmetries already used to define
$\mathcal H_{\rm sym}$.  Simultaneous spectral decomposition of the
relevant commuting invariants therefore induces a further
decomposition
\begin{equation}
   \mathcal H_{\rm sym}
   =
   \bigoplus_a \mathcal H_a,
   \qquad
   [P_a,G]=0
   \quad
   \forall\,G\in\mathfrak g,
   \label{eq:residual_decomposition}
\end{equation}
where $P_a$ projects onto the residual invariant block
$\mathcal H_a$.  Because every element of $e^{\mathfrak g}$ commutes
with $P_a$, the corresponding block populations
$p_a(\psi)=\langle\psi|P_a|\psi\rangle$ are conserved throughout the
entire variational orbit.  In particular,
$p_a(U\psi_{\rm in})=p_a(\psi_{\rm in})$ for all
$U\in e^{\mathfrak g}$.  The circuit can therefore explore states
within each invariant block while remaining unable to redistribute
probability between different blocks.

This makes it important to distinguish \emph{physical symmetries} from
\emph{residual ansatz symmetries}.  Physical symmetries, such as fixed
total magnetization or conserved spin-resolved particle numbers,
define the sector in which the many-body problem is intentionally
solved and should therefore be preserved by the circuit.  Residual
ansatz symmetries instead arise from the chosen generator set and its
connectivity and need not correspond to symmetries of the target
Hamiltonian.  They can consequently fragment an already
symmetry-resolved Hilbert space into smaller invariant components and
exclude physically relevant states even though all intended conserved
quantities are respected.  A useful circuit modification should
therefore preserve the physical symmetry while breaking an unwanted
residual one, schematically satisfying
$[K,Q_{\rm physical}]=0$ but
$[K,Q_{\rm residual}]\neq0$.

\subsection{Residual-population fidelity ceiling}
\label{subsec:block_fidelity_bound}

The invariant block decomposition provides a direct quantitative
measure of the accessibility restriction.  Let the initial and target
states be written as
$\ket{\psi_{\rm in}}
=\bigoplus_a\sqrt{p_a^{\rm in}}\ket{\phi_a^{\rm in}}$
and
$\ket{\psi_0}
=\bigoplus_a\sqrt{p_a^{(0)}}\ket{\phi_a^{(0)}}$,
where the block states are normalized within $\mathcal H_a$.  Since
the residual dynamics preserves the populations $p_a^{\rm in}$, every
state on the accessible orbit can be written as
$\ket{\psi}
=\bigoplus_a\sqrt{p_a^{\rm in}}\ket{\phi_a}$.
Maximizing the overlap independently inside each residual block gives
$|\langle\psi_0|\psi\rangle|
\le
\sum_a\sqrt{p_a^{\rm in}p_a^{(0)}}$, and therefore
\begin{equation}
   F
   \le
   F_{\rm block}
   =
   \left(
   \sum_a
   \sqrt{p_a^{\rm in}p_a^{(0)}}
   \right)^2.
   \label{eq:block_fidelity_ceiling}
\end{equation}
The bound depends only on the mismatch between the residual-block
populations of the input and target states.  If
$p_a^{\rm in}=p_a^{(0)}$ for all $a$, the residual invariant does not
by itself prevent unit fidelity.  If the populations differ, however,
no unitary generated by $\mathfrak g$ can reproduce the target
distribution among the residual blocks, and the maximum accessible
fidelity is strictly smaller than one.  This ceiling is therefore
kinematic rather than optimization-induced: increasing circuit depth
may improve the representation of states within each block, but it
cannot transfer population between blocks or remove the resulting
fidelity restriction.

The bound should not be interpreted as a statement that an optimized
VQE state must saturate $F_{\rm block}$.  The state that maximizes the
target overlap subject to fixed residual populations is generally
different from the state that minimizes the energy on the same
restricted orbit.  If $\ket{\psi_{\rm ov}}$ denotes the
maximum-overlap state and $\ket{\psi_{\rm E}}$ the energy-minimizing
state, one can in general have
\[
   |\langle\psi_0|\psi_{\rm E}\rangle|^2
   <
   |\langle\psi_0|\psi_{\rm ov}\rangle|^2
   \le
   F_{\rm block}.
\]
The residual ceiling therefore characterizes the best fidelity
compatible with the algebraic invariants, while the actual VQE
performance additionally depends on finite-depth expressibility and
optimization.  This distinction motivates the constructive problem
addressed in the following sections: once the residual invariant
responsible for the ceiling has been identified, which physically
allowed generator should be added to couple the corresponding blocks
and enlarge the accessible orbit?

\section{Residual Escape and Algebraic Completion}
\label{sec:escape_theorems}

\subsection{Real Two-Block Escape and Orthogonal Completion}
\label{subsec:real_escape}

We first consider the simplest nontrivial residual decomposition, $V=V_1\oplus V_2$, with $\dim V_1=p$ and $\dim V_2=q$, and assume that the residual algebra acts independently within the two blocks as $\mathfrak r=\mathfrak{so}(p)\oplus\mathfrak{so}(q)$. In a basis adapted to this decomposition, residual generators have the form $R(A,B)=\operatorname{diag}(A,B)$, with $A\in\mathfrak{so}(p)$ and $B\in\mathfrak{so}(q)$. A physical generator that couples the two residual sectors contributes an off-diagonal component
\begin{equation}
   X_C=
   \begin{pmatrix}
      0&C\\
      -C^T&0
   \end{pmatrix},
   \qquad
   C\in M_{p\times q}(\mathbb R),
   \label{eq:real_interblock_generator}
\end{equation}
where $C=P_1KP_2$ is the inter-block part of the physical generator $K$ in the residual basis. The full space of such real inter-block directions is $\mathcal M_{\mathbb R}=\{X_C:C\in M_{p\times q}(\mathbb R)\}$ and has real dimension $pq$. As a vector space, $\mathfrak{so}(p+q)=[\mathfrak{so}(p)\oplus\mathfrak{so}(q)]\oplus\mathcal M_{\mathbb R}$.

\begin{figure}[t]
   \centering
   \includegraphics[width=\columnwidth]{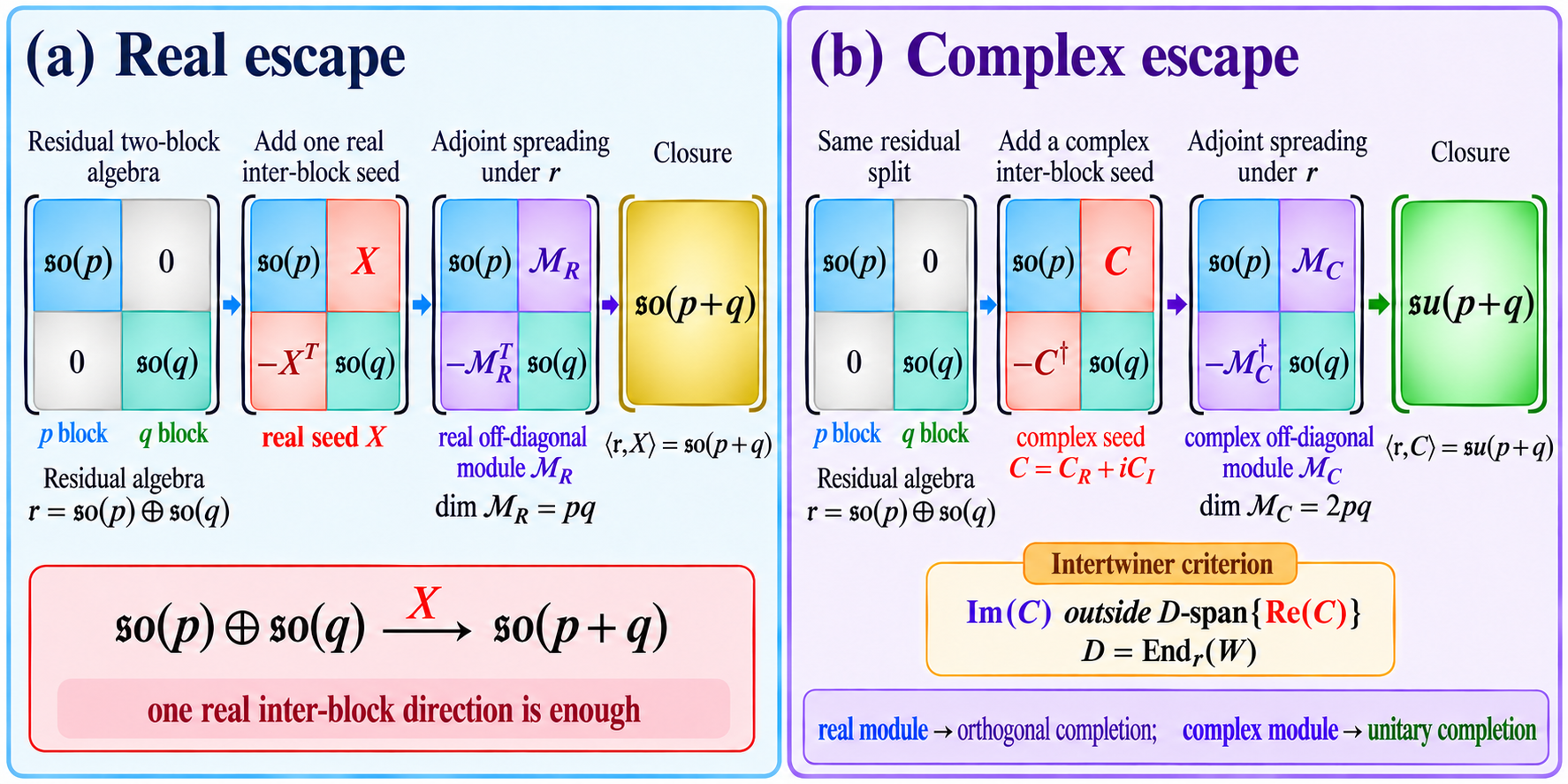}
   \caption{Schematic representation of the two-block escape theorems. For a residual algebra $\mathfrak r=\mathfrak{so}(p)\oplus\mathfrak{so}(q)$, a nonzero real inter-block seed spreads under the residual adjoint action to the full real coupling module of dimension $pq$, yielding the orthogonal completion $\mathfrak{so}(p+q)$. For a complex seed $C=C_R+iC_I$, independence of $C_R$ and $C_I$ over the residual intertwiner algebra produces the full $2pq$-dimensional complex coupling module and the special-unitary completion $\mathfrak{su}(p+q)$.}
   \label{fig:two-block_theorem}
\end{figure}


\begin{theorem}[{\bf Two-block orthogonal escape}]
\label{thm:two_block_real}
Let
\[
   \mathfrak r
   =
   \mathfrak{so}(p)\oplus\mathfrak{so}(q)
   \subset
   \mathfrak{so}(p+q),
\]
and let
\[
   X_{ij}=
   \begin{pmatrix}
      0&e_ie_j^T\\
      -e_je_i^T&0
   \end{pmatrix}
\]
be any nonzero elementary inter-block rotation, where $e_i$ and $e_j$
denote the corresponding standard basis vectors. Then
\begin{equation*}
   \operatorname{Lie}\!\left(
   \mathfrak{so}(p)\oplus\mathfrak{so}(q),X_{ij}
   \right)
   =
   \mathfrak{so}(p+q).
   \label{eq:real_escape_theorem}
\end{equation*}
\end{theorem}

\begin{proof}
The residual algebra acts on the inter-block matrix $C$ according to
\[
   [R(A,B),X_C]=X_{AC-CB},
\]
so repeated commutators generate the adjoint module of $C$ under
$C\mapsto AC-CB$.

Without loss of generality, take the elementary seed
$C_{11}=e_1e_1^T$.  Let $E_{ab}=e_ae_b^T$ denote the standard matrix
unit of the appropriate block dimension.  Left multiplication by
\[
   A_{i1}=E_{i1}-E_{1i}\in\mathfrak{so}(p)
\]
generates $e_ie_1^T$, while the right action of
\[
   B_{j1}=E_{j1}-E_{1j}\in\mathfrak{so}(q)
\]
subsequently generates $e_ie_j^T$.  Since the matrices $e_ie_j^T$
form a basis of $M_{p\times q}(\mathbb R)$, the adjoint module
generated from $X_{11}$ is the full real inter-block space
\[
   \mathcal M_{\mathbb R}
   =
   \left\{
   \begin{pmatrix}
      0&C\\
      -C^T&0
   \end{pmatrix}
   :
   C\in M_{p\times q}(\mathbb R)
   \right\},
   \qquad
   \dim\mathcal M_{\mathbb R}=pq.
\]

The residual algebra already contains all block-diagonal
skew-symmetric directions, and
\[
   \mathfrak{so}(p+q)
   =
   \bigl[
   \mathfrak{so}(p)\oplus\mathfrak{so}(q)
   \bigr]
   \oplus
   \mathcal M_{\mathbb R}
\]
as a vector-space decomposition.  Hence the generated Lie algebra
contains every direction in $\mathfrak{so}(p+q)$.  Conversely, all
initial generators are real skew-symmetric, and this property is
preserved under commutation.  Therefore
\[
   \operatorname{Lie}\!\left(
   \mathfrak{so}(p)\oplus\mathfrak{so}(q),X_{ij}
   \right)
   =
   \mathfrak{so}(p+q).
\]
\end{proof}

The theorem gives a direct circuit-design interpretation. If two residual blocks are already internally connected by $\mathfrak{so}(p)$ and $\mathfrak{so}(q)$ dynamics, then a single physical generator with a suitable bridge between them can spread into the full set of inter-block rotations and remove the residual separation completely.

\subsection{Complex Two-Block Escape and Unitary Completion}
\label{subsec:complex_escape}

A physical SP generator can acquire genuinely complex
matrix elements when expressed in the residual basis.  We therefore
consider an inter-block generator of the form
\begin{equation*}
   X_C=
   \begin{pmatrix}
      0&C\\
      -C^\dagger&0
   \end{pmatrix},
   \qquad
   C\in M_{p\times q}(\mathbb C),
   \label{eq:complex_interblock_generator}
\end{equation*}
and write $C=C_R+iC_I$, with
$C_R,C_I\in W\equiv M_{p\times q}(\mathbb R)$.  The residual algebra
$\mathfrak r=\mathfrak{so}(p)\oplus\mathfrak{so}(q)$ acts through
$C\mapsto AC-CB$, so the real and imaginary parts transform
independently under the same real representation.  Their relevant
notion of independence is therefore determined by the intertwiner
algebra
\[
   \mathbb D=\operatorname{End}_{\mathfrak r}(W),
\]
consisting of real linear maps on $W$ that commute with the residual
action.

This produces a qualitative distinction from the real case.  A complex
seed can generate two independent copies of the real coupling module,
giving $2pq$ real inter-block directions.  Whether both copies are
generated is controlled not by ordinary real linear independence of
$C_R$ and $C_I$, but by their independence over $\mathbb D$.  Once the
full complex inter-block module is generated, its self-commutators
supply the additional block-diagonal anti-Hermitian directions required
for special-unitary completion.

\begin{theorem}[{\bf Two-block complex escape and unitary completion}]
\label{thm:complex_escape}
Let $\mathfrak r=\mathfrak{so}(p)\oplus\mathfrak{so}(q)\subset\mathfrak{su}(p+q)$ act irreducibly on $W=M_{p\times q}(\mathbb R)$ by $(A,B)\cdot C=AC-CB$, and let $\mathbb D=\operatorname{End}_{\mathfrak r}(W)$.  If $C=C_R+iC_I$ with $C_R$ and $C_I$ linearly independent over $\mathbb D$, then
\begin{equation*}
   \mathcal M_C=\{X_Z:Z\in M_{p\times q}(\mathbb C)\},
   \qquad
   \dim_{\mathbb R}\mathcal M_C=2pq,
   \label{eq:full_complex_escape}
\end{equation*}
and
\begin{equation*}
   \operatorname{Lie}\!\left(
   \mathfrak{so}(p)\oplus\mathfrak{so}(q),X_C
   \right)=\mathfrak{su}(p+q).
   \label{eq:complex_completion}
\end{equation*}
\end{theorem}

\begin{proof}
Since the residual representation on $W$ is irreducible, the real
density theorem implies that the associative algebra generated by
$\mathfrak r$ acts as
$\operatorname{End}_{\mathbb D}(W)$.  Identifying the complex coupling
space with $W\oplus W$, the residual action on the seed is therefore
\[
   T\cdot(C_R,C_I)
   =
   (TC_R,TC_I),
   \qquad
   T\in\operatorname{End}_{\mathbb D}(W).
\]
Because $C_R$ and $C_I$ are independent over $\mathbb D$, they may be
extended to a $\mathbb D$-basis of $W$.  Hence, for arbitrary
$X,Y\in W$, there exists a $\mathbb D$-linear endomorphism $T$ such
that $TC_R=X$ and $TC_I=Y$.  Repeated residual commutators therefore
generate every pair $(X,Y)\in W\oplus W$, or equivalently every
$Z=X+iY\in M_{p\times q}(\mathbb C)$.  Thus
$\mathcal M_C$ is the full complex inter-block module and has real
dimension $2pq$.

It remains to determine the algebra generated once this module is
available.  For any $X_C,X_D\in\mathcal M_C$,
\[
   [X_C,X_D]
   =
   \begin{pmatrix}
      -CD^\dagger+DC^\dagger&0\\
      0&-C^\dagger D+D^\dagger C
   \end{pmatrix}.
\]
As $C$ and $D$ range over all complex $p\times q$ matrices, these
commutators generate
\[
   \mathfrak{s}\!\left(
      \mathfrak u(p)\oplus\mathfrak u(q)
   \right),
\]
which has real dimension $p^2+q^2-1$.  Together with the $2pq$
off-diagonal directions,
\[
   (p^2+q^2-1)+2pq
   =
   (p+q)^2-1
   =
   \dim\mathfrak{su}(p+q).
\]
All starting generators are traceless and anti-Hermitian, and these
properties are preserved under commutation.  Hence the generated Lie
algebra is exactly $\mathfrak{su}(p+q)$.
\end{proof}

If $C_I$ instead lies in the $\mathbb D$-span of $C_R$, only one copy
of $W$ is generated and the inter-block module has dimension $pq$.  In
the generic real-type case, $\mathbb D=\mathbb R I$, the full-complex
condition therefore reduces simply to $C_I\not\propto C_R$.  When one
block has dimension two, the intertwiner algebra is enlarged by the
invariant rotation
\[
   J=
   \begin{pmatrix}
      0&-1\\
      1&0
   \end{pmatrix}.
\]
For $q=2$, full complex escape requires
$C_I\notin\operatorname{span}_{\mathbb R}\{C_R,C_RJ\}$, while for
$p=2$ the analogous condition is
$C_I\notin\operatorname{span}_{\mathbb R}\{C_R,JC_R\}$.
The case $p=q=2$ is exceptional and is not covered by the irreducibility
assumption of the theorem.

\begin{figure}[t]
   \centering
   \includegraphics[width=\columnwidth]{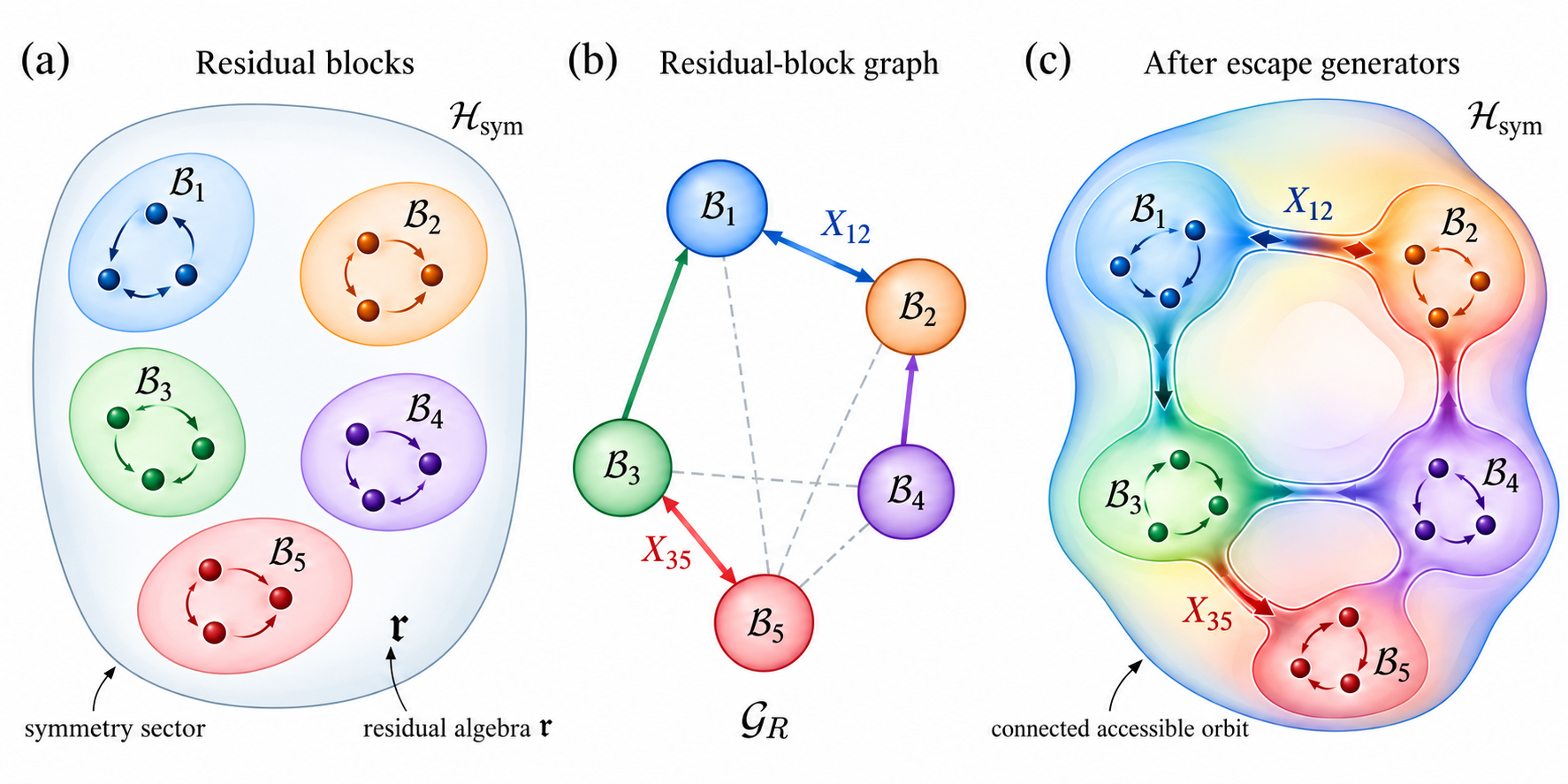}
   \caption{Residual-block representation of algebraic escape. (a) The residual algebra $\mathfrak r$ decomposes the physical symmetry sector into invariant blocks $\mathcal B_1,\ldots,\mathcal B_5$, within which the baseline generators produce independent internal dynamics. (b) Candidate physical escape generators appear as edges of the residual-block graph $\mathcal G_R$; examples $X_{12}$ and $X_{35}$ directly couple distinct invariant sectors. (c) When the selected escape edges connect the residual-block graph, the previously disconnected blockwise orbits merge into a larger connected accessible orbit inside the same physical symmetry sector.}
   \label{fig:multi-block_theorem}
\end{figure}

\subsection{Multi-Block Escape and Residual-Graph Completion}
\label{subsec:multiblock_escape}

The two-block escape theorems extend naturally when the residual commutant decomposes the physical symmetry sector into more than two invariant components. We consider $V=\bigoplus_{a=1}^{m}V_a$, with $\dim V_a=d_a$, and residual algebra $\mathfrak r=\bigoplus_{a=1}^{m}\mathfrak{so}(d_a)$. A physical generator $K$ is resolved into block components $P_aKP_b$, where diagonal terms act within existing residual sectors and off-diagonal terms provide possible escape channels. This structure defines a residual-block graph $\mathcal G_{\rm res}=(\mathcal V,\mathcal E)$ whose vertices are the invariant blocks and whose edges correspond to selected generators satisfying $P_aKP_b\neq0$. The graph describes connectivity between residual many-body sectors rather than connectivity between physical lattice sites. For the real escape problem, an edge is complete when its seed generates the full real coupling module $M_{d_a\times d_b}(\mathbb R)$.

\begin{corollary}[{\bf Connected real residual graph}]
\label{cor:connected_real_graph}
Let $\mathfrak r=\bigoplus_{a=1}^{m}\mathfrak{so}(d_a)$, and let a selected set of physical generators define a connected residual-block graph. Assume that every selected edge $(a,b)$ generates the full compatible real coupling module between $V_a$ and $V_b$. Then
\begin{equation}
   \operatorname{Lie}\!\left(
   \bigoplus_{a=1}^{m}\mathfrak{so}(d_a),\{K_e\}
   \right)
   =
   \mathfrak{so}\!\left(\sum_{a=1}^{m}d_a\right).
   \label{eq:multiblock_real_completion}
\end{equation}
\end{corollary}

\begin{proof}
Choose any spanning tree of the connected residual-block graph. Apply Theorem~\ref{thm:two_block_real} to one tree edge to merge its incident blocks. Treat the merged block as a single vertex and repeat along the remaining tree edges. Since a spanning tree on $m$ vertices contains $m-1$ edges, after $m-1$ mergers the resulting algebra is $\mathfrak{so}(\sum_a d_a)$.
\end{proof}
The graph formulation converts minimal circuit completion into a graph optimization problem. In the clean real case, a fully connected physical coupling pattern is unnecessary: any spanning tree of valid escape edges is sufficient. If each physically available edge is assigned a cost reflecting gate count, hardware distance, implementation error, or locality, one may therefore search for a minimum-cost spanning tree subject to the requirement that every chosen edge generates the required module.

\section{Pre-VQE Residual-Algebra Circuit Design}
\label{sec:circuit_design}
\begin{figure}[t]
   \centering
   \includegraphics[width=\columnwidth]{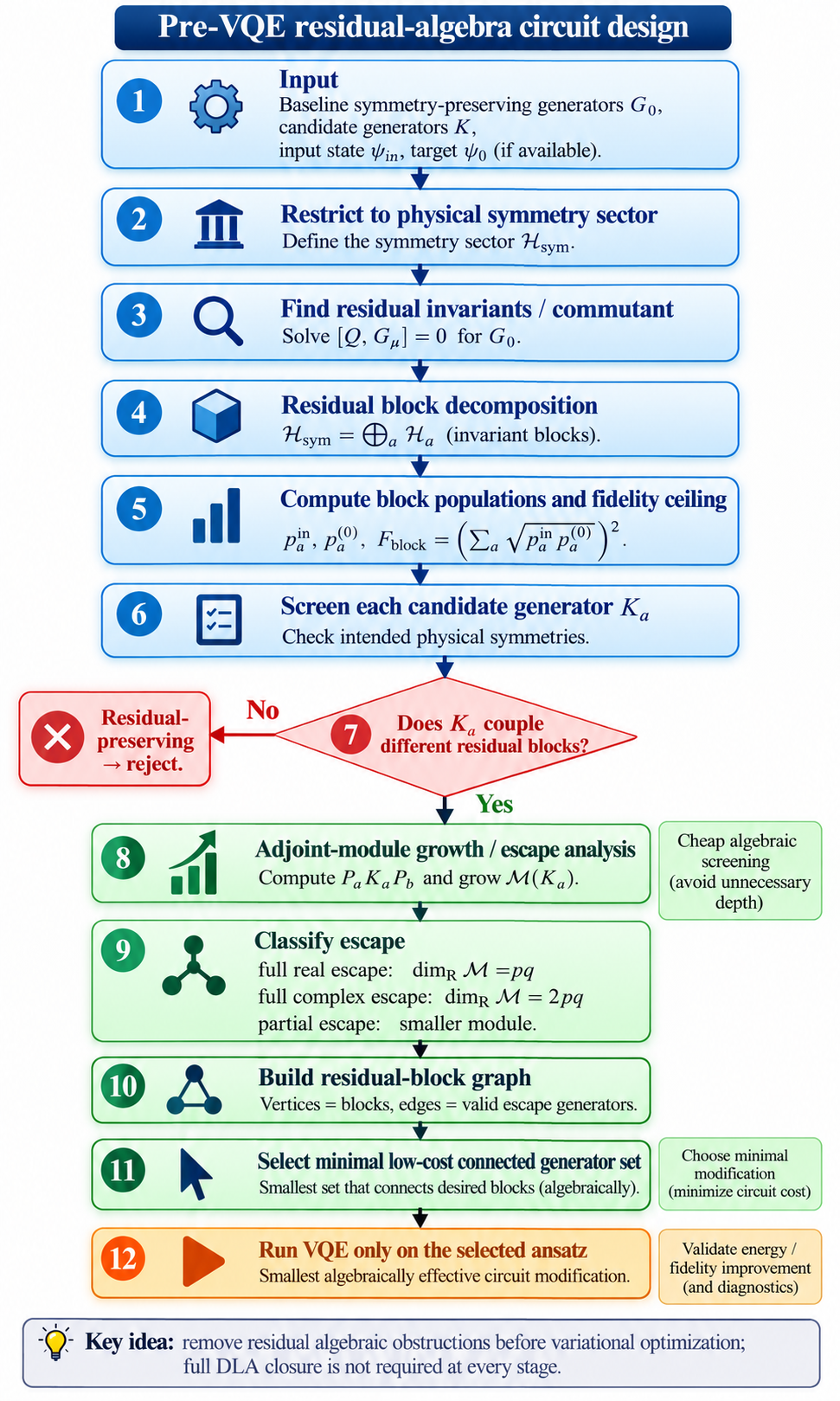}
   \caption{
   Pre-VQE residual-algebra circuit-design workflow.  Starting from a
   baseline SP generator set, residual invariants and
   invariant-block populations are identified first. Candidate physical
   generators are then screened through their inter-block couplings and
   adjoint-module growth.  The smallest set predicted to remove the
   algebraic obstruction is selected before finite-depth VQE is performed.
   }
   \label{fig:circuit_design}
\end{figure}

The preceding results suggest a circuit-design strategy in which candidate generators are screened algebraically before expensive variational optimization. The workflow is shown in Fig.\ref{fig:circuit_design}. Starting from a baseline SP generator set $\mathcal G_0=\{G_\mu\}$, the first objective is to identify residual constraints that restrict the accessible orbit inside the intended physical symmetry sector $\mathcal H_{\rm sym}$. These constraints are obtained from the commutant of the baseline algebra, $\mathrm{Comm}(\mathfrak g_0)=\{Q:[Q,G_\mu]=0,\ \forall\mu\}$. Numerically, we identify this commutant through the positive-semidefinite superoperator $\mathcal K(X)=\sum_\mu[H_\mu,[H_\mu,X]]$, where $H_\mu=iG_\mu$ and $X$ is an arbitrary operator on $\mathcal H_{\rm sym}$. Its zero modes satisfy $\mathcal K(X)=0$ and are precisely the operators that commute with all baseline generators, thereby providing the nontrivial residual invariants. Nontrivial Hermitian zero modes define conserved operators whose spectral projectors $P_a$ decompose the physical symmetry sector as: $\mathcal H_{\rm sym}=\bigoplus_a\mathcal H_a$
which readily imposes the fidelity bound to the maximum $F_{\rm block}$ idependent of circut depth (see Eq.~\eqref{eq:block_fidelity_ceiling} and \ref{eq:residual_decomposition} for details). 



To break this residual obstruction, candidate physical generators $K_\alpha$ are then screened through their
residual-block components $C_{ab}=P_aK_\alpha P_b$. A generator satisfying
$C_{ab}=0$ for all $a\neq b$ preserves the residual decomposition
and cannot remove a population mismatch.  Conversely, $C_{ab}\ne 0$ identify possible escape channels.  Their algebraic effect is
determined from the adjoint module generated under the residual algebra,
\[
   \mathcal M(K_\alpha)
   =
   {\rm span}
   \left\{
   K_\alpha,\,
   [\mathfrak g_{\rm res},K_\alpha],\,
   [\mathfrak g_{\rm res},
   [\mathfrak g_{\rm res},K_\alpha]],\ldots
   \right\}.
\]

For a two-block decomposition of dimensions $p$ and $q$, a full real
module of dimension $pq$ predicts orthogonal completion, whereas a full
complex module of real dimension $2pq$ predicts special-unitary completion.
Candidates producing smaller modules represent partial escape.  For
multiple residual blocks, the surviving candidates define a residual-block
graph, allowing a minimal or minimum-cost connected set of escape generators
to be selected.

VQE enters only in the final stage.  Its purpose is not to discover which
physical generator should be added, but to determine how efficiently the
algebraically selected circuit realizes the enlarged accessible manifold at
finite depth.  The XXZ and Hubbard models below provide complementary
worked examples of this procedure: the XXZ model realizes a real
Hodge-sector escape mechanism, whereas the Hubbard model realizes a complex
inter-block escape and special-unitary completion.

\begin{figure*}[t]
   \centering
   \includegraphics[width=0.99\textwidth]{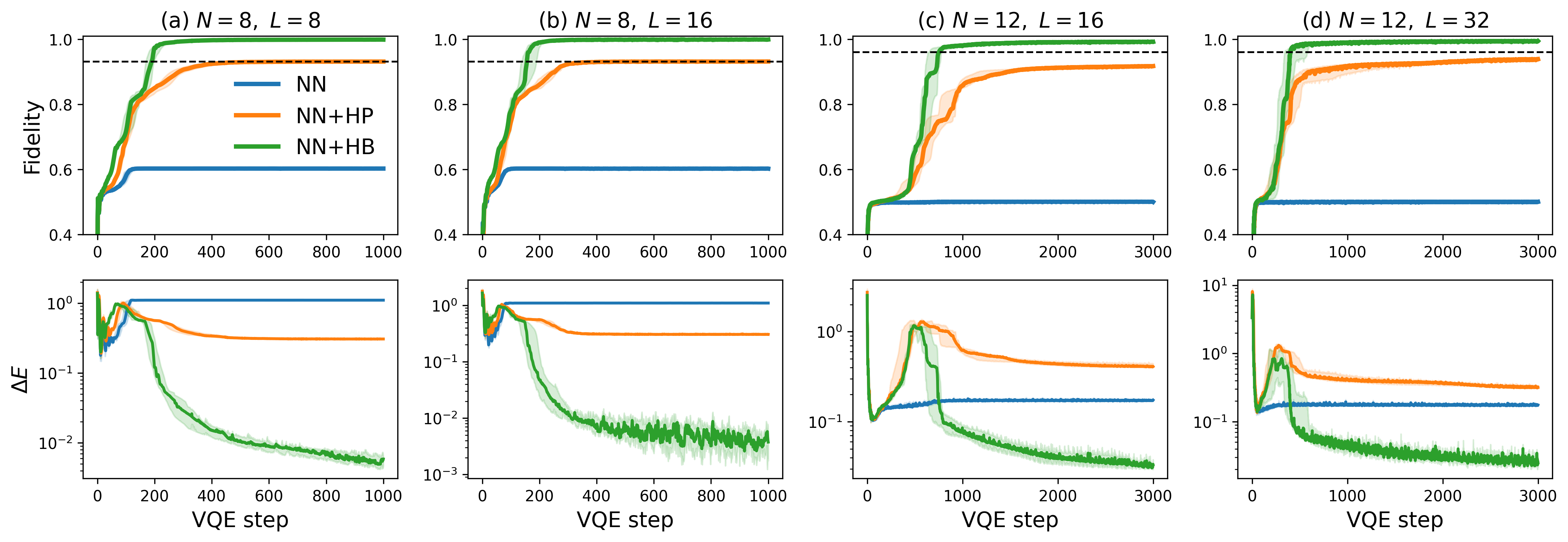}
   \caption{Finite-depth VQE verification for the $S_z^{tot}=0$ XXZ model at $J_z=2.5$. Top panel: (a), (b) show the fidelity vs the VQE steps for $N=8$ and circuit depth $L=8, 16$ respectively. The green, orange and blue colors are for NN, Hodge preserving and Hodge breaking ans\"atze respectively as indicated. The dashed line represents the fidelity bound set by residual obstrcution. The smimilar fidelity plot for large system $N=12$ and circuit depth $L=16, 32$ respectively are shown in (a) and (b). The corresponding $\Delta E$ are depicted in lower panel. Solid curves show the median over different random-seed runs, with shaded interquartile range.}
   \label{fig:xxz_vqe}
\end{figure*}

\section{Worked Example I: XXZ model}
\label{sec:xxz_example}

\subsection{Residual obstruction and its breaking}
\label{subsec:xxz_residual}

We first apply the pre-VQE residual-algebra design procedure to the periodic spin-$1/2$ XXZ model, $H_{\rm XXZ}=\sum_j(S_j^xS_{j+1}^x+S_j^yS_{j+1}^y+J_zS_j^zS_{j+1}^z)$. Conservation of total magnetization restricts the dynamics to a fixed-excitation sector carrying the exterior-power representation $\mathcal H_{N,k}\simeq\Lambda^k(\mathbb R^N)$ of Hilbert space dimension $D=\binom Nk$. The baseline SP circuit is generated by nearest-neighbor (NN) single-excitation (SE) operators and realizes the induced algebra $\mathfrak g_{\rm NN}=\rho_k[\mathfrak{so}(N)]$.

To demonstrate the residual-algebra design procedure explicitly, we first consider the relatively small $N=8$ ring with $S_z^{tot}=0$, corresponding to $k=N/2$ and $D=70$. At half-filling the Hodge star operator $\star:\Lambda^k(\mathbb R^{2k})\rightarrow\Lambda^k(\mathbb R^{2k})$ acts within the physical symmetry sector and satisfies $\star^2=I$. Since the NN exterior-power algebra commutes with $\star$, the symmetry sector decomposes into two eigenspaces: $\mathcal H_{\rm sym}=\mathcal H_+\oplus\mathcal H_-$ of equal dimension $\dim\mathcal H_\pm=35$. This residual block decomposition restrict the dynamics of NN circuts within each block, yeilding the fidelity ceiling depending on initial and target states. Furthermode, the NN circuit algebra has only $\dim\mathfrak g_{\rm NN}=28$, much smaller than either $\mathfrak{so}(35)$ block and vastly smaller than the full $\mathfrak{so}(70)$ algebra of the symmetry sector. The baseline NN SE circuit is therefore restricted not only by the Hodge-sector populations, but also by incomplete controllability within each residual block.

We next ask which additional exchange generators can escape this restricted algebra. For a candidate generator $K=K_\parallel+K_\perp$, with $K_\parallel\in\mathfrak g_{\rm NN}$, we define the algebraic escape fraction $\eta_{\rm out}(K)=\|K_\perp\|/\|K\|$. For the relevant nonlocal exchange class in the exterior-power representation, we derived an analytical expression: $\eta_{\rm out}=2\sqrt{(k-1)(N-k-1)}/(N-2)$ which becomes nonzero at finite $k$ and grows toward intermediate filling, revealing that the additional directions generated by nonlocal connectivity are intrinsically many-body. Algebraic escape alone is insufficient; a candidate breaks the Hodge obstruction only if it has nonzero inter-sector matrix elements, $P_\pm K P_\mp\neq0$.

The distinction is illustrated by the additional SE generator $K_{03}$ acting on $(0,3)$ pair of qubits. This resulting circuit NN+$(0,3)$ is Hodge preserving (HP): $[\star,K_{03}]=0$, and therefore has no inter-block component, i.e, $P_\pm K_{03}P_\mp=0$. Nevertheless, its adjoint closure strongly enlarges the algebra within the two Hodge sectors, completing each block separately:
$\mathfrak g_{\rm HP}\simeq
\mathfrak{so}(35)\oplus\mathfrak{so}(35); \dim\mathfrak g_{\rm HP}=1190.$
Thus the HP connection removes the intra-block restriction of the NN circuit but leaves the residual decomposition intact. The Hodge populations remain conserved, and the fidelity ceiling $F_{block}$ therefore survives irrespective of circuit depth.

By contrast, the $(0,2)$ SE connection is Hodge breaking (HB) as it anticommutes with the Hodge star: $\{\star,K_{02}\}=0$. It consequently acts off-diagonally with respect to the Hodge decomposition: $P_\pm K_{02}P_\pm=0$, while $P_\pm K_{02}P_\mp\neq0$. The added generator therefore directly couples the two residual sectors and breaks the fidelity constratnt. Repeated commutators with the residual algebra propagate this inter-block coupling throughout the sector and complete the algebra in accordance with the Theorem.~\ref{thm:two_block_real} to
$\mathfrak g_{\rm HB}
\simeq
\mathfrak{so}(70); \dim\mathfrak g_{\rm HB}=2415.$
This connectivity sensitivity is consistent with earlier observations that adding selected non-nearest-neighbor excitation links can strongly enlarge variational accessibility \cite{Reja2026Connectivity}; here we resolve the mechanism through the residual Hodge structure.

Having established the full algebraic mechanism for $N=8$, we then consider the larger half-filled $N=12$ ring, with $D=924$, using only the scalable diagnostics of the framework rather than full Lie closure. The Hodge commutation test and residual block components $P_aKP_b$ again classify $(0,3)$ as Hodge preserving and $(0,2)$ as Hodge breaking. Thus the same screening principle remains applicable at larger system size.

\subsection{Finite-depth VQE verification}
\label{subsec:xxz_vqe}
The finite-depth VQE calculations are implemented in PennyLane using the \texttt{lightning.qubit} backend and optimized with the Adam optimizer. All runs start from the same half-filled alternating product state, $\ket{\psi_{\rm in}}=\ket{0101\cdots}$, in the $S_z^{\rm tot}=0$ sector, while only the initial variational parameters are randomized across seeds. To test the algebraic predictions above, we run the simulation for NN, NN+HP, and NN+HB circuits for both $N=8$ and $N=12$ XXZ rings at multiple circuit depths and over several seeds. We monitor the fidelity and energy error trajectory vs the VQE steps which are shown in Fig.~\ref{fig:xxz_vqe}. 
As indicated, the NN, NN+$U$, and NN+$D$ ans\"atze results for fidleity and $\Delta E$ are shown in blue, orange, and green lines, respectively. For each circuit, the solid lines represents the median over random-seed runs and the shaded region the interquartile range. The dashed line denotes the residual block-population fidelity ceiling. 

As predicted in Fig.~\ref{fig:xxz_vqe}, the VQE results follow the residual algebra analyses. The NN circuit for $N=8, L=8$ XXZ ring in (a), fidelity remains strongly restricted at sufficiently large VQE steps. The NN+HP circuit quickly improves the fidelity substantially, and saturates to the fidelity ceiling (dashed line) set by the residual obstruction. The NN+HB circuit, however, breaks the fidelity ceiling and quickly saturates to unit fidelity. Larger depth $L=16$ results for NN and NN+HP circuits shown in (b) demonstrate that the residual ostruction and corresponding fidleity ceiling can not be broken by increasing circuit depth. The corresponding energy errors for NN and NN+HP saturate at relatively large values whereas error gets lowered by several order of magnitude for NN+HB circuit.

The results equally holds for larger XXZ $N=12$ ring as depicted in Fig.~\ref{fig:xxz_vqe}(c) and (d). Increasing circuit depth from $L=16$ to $32$ give small quantative improvement for NN+HP circuits, but it can not break the residual symmetry and remains below the fidleity ceiling (dashed line) even at sufficiently large VQE steps; while NN+HB circuit breaks this obstruction and attains the maximum fidelity. The corresponding energy error gets lowest in NN+HB circuit as compared to restricted NN and NN+HP circuits.       

\begin{figure*}[t]
   \centering
   \includegraphics[width=0.99\textwidth]{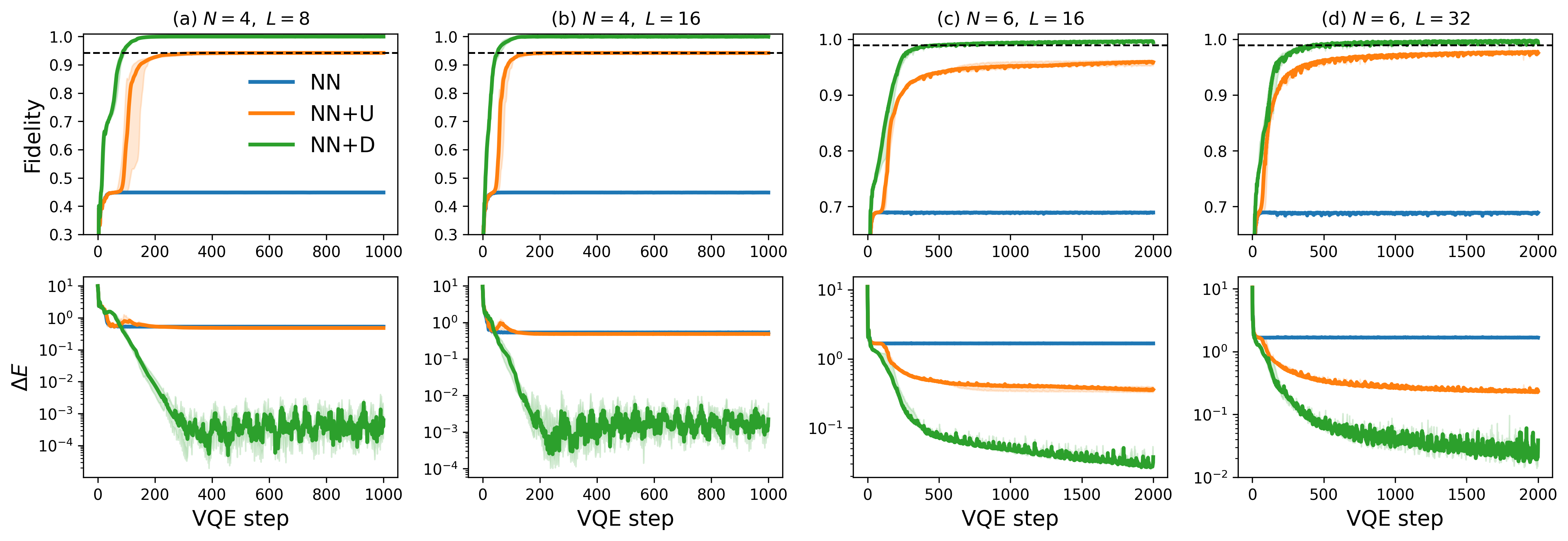}
   \caption{Finite-depth VQE results for the half-filled Hubbard model at $U/t=4$. Top panel: the fidelity vs VQE steps for (a) $N=4$, $L=8$; (b) $N=4$, $L=16$; (c) $N=6$, $L=16$; and (d) $N=6$, $L=32$. The green, orange and blue colors are for NN, NN+$U$ and NN+$D$ ans\"atze respectively as indicated. The dashed horizontal line marks the residual block-population fidelity ceiling. Bottom panel: depicts the corresponding energy errors $\Delta E=E_{\rm VQE}-E_0$. Solid curves show the median over different random-seed runs, with shaded interquartile ranges.}
   \label{fig:hubbard_vqe}
\end{figure*}

\section{Worked Example II: Hubbard Ring}
\label{sec:hubbard_example}
\subsection{Residual obstruction and its breaking}
\label{subsec:hubbard_residual}

We next apply the residual-algebra design procedure to the half-filled Hubbard ring with $S_z^{tot}=0$. The baseline SP circuit is generated by NN fermionic SE operators $G_{ij,\sigma}=c_{i\sigma}^{\dagger}c_{j\sigma}-c_{j\sigma}^{\dagger}c_{i\sigma}$ for $\sigma=\uparrow,\downarrow$, and therefore preserves $N_\uparrow$ and $N_\downarrow$ exactly.
In general, for the half-filled NN hopping algebra, the separate middle-exterior-power structure of the two spin species gives a four-dimensional commutant, generated by the identity and the two spin-resolved Hodge operators and their product. 

Again, we first choose a half-filled \(N=4\) system to present the working framework of our circuit design principle. The NN circuit generates a restricted dynamical algebra $\mathfrak g_{\rm NN}\simeq\mathfrak{so}(4)_\uparrow\oplus\mathfrak{so}(4)_\downarrow$ with $\dim\mathfrak g_{\rm NN}=12$ and a four-dimensional commutant, revealing residual constraints beyond the imposed particle-number symmetries. 

To break the residual obstructions, we first enlarge the circuit by adding the Hamiltonian-inspired onsite interaction generator $G_U=-i\sum_i n_{i\uparrow}n_{i\downarrow}$ with single variational parameter per layer. The resulting NN+$U$ circuit has residual algebra $\mathfrak g_{\rm NN+U}\simeq\mathfrak{so}(18)\oplus\mathfrak{so}(18)$, while its commutant is reduced to the identity and a single nontrivial Hermitian involution $Q$, satisfying $Q^2=I$.  The projectors $P_\pm=(I\pm Q)/2$ therefore decompose the $D=36$ symmetry sector into two $18$-dimensional invariant subspaces.  Thus the onsite interaction removes part of the NN residual structure but leaves a final two-block obstruction that prevents population transfer between the sectors, yielding the circuit depth-independent fidelity bound.

We then test a correlated doublon-transfer generator,
$G_D=c_{j\uparrow}^{\dagger}c_{j\downarrow}^{\dagger}c_{i\downarrow}c_{i\uparrow}-\mathrm{H.c.}$, against this surviving decomposition. For the resulting NN+$D$ circuit, we find that it anticommutes with the residual involution: $\{Q,G_D\}\simeq0$, so its intra-block components vanish, $P_\pm G_dP_\pm\simeq0$, while $P_\pm G_dP_\mp\neq0$. The generator therefore acts directly between the two residual sectors and breaks the conserved-population obstruction. Repeated adjoint action of $\mathfrak g_{\rm NN+U}$ on $G_D$ generates the full complex inter-block module of real dimension $2(18)^2=648$. Together with the induced block-diagonal directions, this completes the full complex algebra in accordance with Theorem.~\ref{thm:complex_escape} as: 
$\mathfrak{so}(18)\oplus\mathfrak{so}(18)
\xrightarrow{\;G_D\;}
\mathfrak{su}(36).$

Having established the full residual-algebra mechanism for $N=4$, we next examine the larger half-filled $N=6$ system with $D=400$. At this size, we focus on the scalable components of the framework—commutant diagnostics, residual block-population constraints, and finite-depth testing of the selected correlated extension. The aim is to determine whether the algebraically identified obstruction continues to predict the relative performance of competing SP circuit architectures as the Hilbert-space dimension increases. The VQE results presented below provide this larger-system validation.

\subsection{Finite-depth VQE verification}
\label{subsec:hubbard_vqe}
The finite-depth VQE are implemented in PennyLane using the \texttt{lightning.qubit} backend and optimized with the Adam optimizer. All the calculations start from the same half-filled Hartree–Fock product state with the first \(N/2\) sites doubly occupied, using the interleaved spin-orbital ordering \((0\uparrow,0\downarrow,1\uparrow,1\downarrow,\ldots)\); only the initial variational parameters are randomized across seeds. To test the algebraic predictions above, we run VQE simulations for the NN, NN+$U$, and NN+$D$ circuits at half filling for both $N=4$ and $N=6$, each for multiple seeds. We monitor the ground-state fidelity $F_{\rm VQE}=|\langle\psi_0|\psi_{\rm VQE}\rangle|^2$ and the energy error $\Delta E=E_{\rm VQE}-E_0$. For the above intitial state, the NN+$U$ circuit conserves the block populations $p_\pm$, which impose the depth-independent fidelity ceiling as $F_{\rm block}=0.94209771$ and $0.98949603$ for $N=4$ and $6$ Hubbard rings respectively at intermediate interaction $U/t=4$.

As depicted in Fig.~\ref{fig:hubbard_vqe}, the numerical results follow the algebraic hierarchy directly.  For $N=4, L=8$ in (a), the NN circuit remains strongly restricted, while NN+$U$ substantially improves the fidelity but saturates at the residual ceiling (dashed line). The NN+$D$ circuit breaks this obstruction, reaches near-unit fidelity. The same fidelity saturation value at $L=16$ in (b) ensures that algebraic obstruction can not be removed by increasing circuit depth. The corresponding energy errors saturate at relatively large values for NN and NN+$U$; while it gets lowered by several additional orders of magnitude for NN+$D$ circuit [see Figs.~\ref{fig:hubbard_vqe}(e,f)].

The same ordering persists for $N=6$.  Increasing the depth from $L=16$ to $L=32$ improves the quantitative convergence, but NN remains poorly expressive and NN+$U$ stays below the predicted fidelity ceiling.  In contrast, NN+$D$ consistently achieves the highest fidelity and the smallest energy error [Figs.~\ref{fig:hubbard_vqe}(c,d,g,h)].  These results confirm that the residual obstruction identified before optimization remains predictive at larger system size, and that breaking the relevant residual blocks is more consequential than simply increasing circuit depth.

\section{Discussion}
\label{sec:discussion}

The residual-algebra framework separates algebraic accessibility, finite-depth expressibility, and optimization efficiency. The DLA and its commutant determine the asymptotic orbit accessible to the circuit; the finite-depth parameterization determines how efficiently that orbit is represented; and the optimizer determines whether a useful point on that manifold is reached. The fidelity ceiling of Eq.~\eqref{eq:block_fidelity_ceiling} belongs to the first level and cannot be removed by increasing depth or changing the optimizer alone. The escape theorems address the same level constructively by identifying which generators modify the residual orbit itself.

A second important distinction is between physical connectivity and residual connectivity. The physical interaction graph specifies which local degrees of freedom are directly coupled, whereas the residual-block graph specifies which invariant many-body sectors are connected after the baseline algebra has been resolved. These graphs need not have the same structure. A local generator can bridge globally defined residual sectors, while an apparently substantial modification of the physical graph can remain algebraically ineffective if its projection between the relevant residual blocks vanishes.

The real and complex escape results further show that the type of the generated module determines the class of the completion algebra. A full real $pq$-dimensional coupling module joins $\mathfrak{so}(p)\oplus\mathfrak{so}(q)$ into $\mathfrak{so}(p+q)$, whereas a full complex $2pq$-dimensional module produces the unitary directions required for $\mathfrak{su}(p+q)$. The intertwiner algebra $\mathbb D=\operatorname{End}_{\mathfrak r}(W)$ provides the natural criterion for distinguishing these cases. Although the present work focuses on real- and complex-type modules, the same representation-theoretic viewpoint suggests a broader classification involving quaternionic intertwiner structure and possible symplectic completion mechanisms.

The XXZ and Hubbard applications demonstrate complementary aspects of the framework. In the $N=8$ half-filled XXZ ring, Hodge-preserving and Hodge-breaking exchange connections both conserve total magnetization, yet only the latter couples the residual Hodge sectors; the same screening distinction is then tested at the larger $N=12$ size using scalable diagnostics. In the half-filled $N=4$ Hubbard system, a correlated doublon-transfer generator is purely inter-block with respect to the residual involution, generates the full $648$-dimensional complex coupling module, and completes $\mathfrak{so}(18)\oplus\mathfrak{so}(18)$ to $\mathfrak{su}(36)$; the corresponding design principle is further examined for $N=6$.

From a computational perspective, explicit construction of the full enlarged DLA is not required at every stage. The residual commutant can often be extracted directly from the original generators, candidate generators can be screened through their block components, and the enlarged algebra can be certified from adjoint-module and commutator-sector dimensions. This is important because exhaustive Lie closure becomes increasingly expensive as the symmetry-sector dimension and generated algebra grow.

The present results are statements about accessibility rather than guaranteed finite-depth performance. A generator that removes the residual obstruction can still require substantial circuit depth before its new directions become well represented in the finite-depth tangent space, and the enlarged parameterization may have an unfavorable optimization geometry. A natural next step is therefore to connect residual adjoint-module growth with finite-depth diagnostics such as tangent rank, quantum Fisher information, and optimization complexity.

\section{Conclusion}
\label{sec:conclusion}

We have developed a residual-algebra framework for diagnosing and removing accessibility restrictions in symmetry-preserving variational quantum circuits. A SP generator set can possess additional residual invariants inside the intended physical symmetry sector, and the corresponding commutant decomposes that sector into invariant blocks whose populations are fixed along the variational orbit. These conserved populations lead directly to the kinematic fidelity ceiling of Eq.~\eqref{eq:block_fidelity_ceiling}.

We then converted this diagnostic into a constructive circuit-design principle. For two residual orthogonal blocks, a single elementary real inter-block connection generates the full real coupling module and completes $\mathfrak{so}(p)\oplus\mathfrak{so}(q)$ to $\mathfrak{so}(p+q)$. For genuinely complex inter-block seeds, independence of the real and imaginary components over the residual intertwiner algebra produces the full complex module and completion to $\mathfrak{su}(p+q)$. The multi-block extension recasts the same mechanism as connectivity in a residual-block graph.

The XXZ and Hubbard applications demonstrate how these algebraic results translate into physical circuit design. In the XXZ model, Hodge-preserving and Hodge-breaking exchange connections can be distinguished directly by their residual block couplings. In the half-filled $N=4$ Hubbard sector, the baseline circuit has residual algebra $\mathfrak{so}(18)\oplus\mathfrak{so}(18)$ and a corresponding two-block fidelity ceiling, while a correlated doublon-transfer generator generates a full complex escape module and completes the algebra to $\mathfrak{su}(36)$.

The resulting strategy is to identify residual invariants, quantify the associated accessibility restriction, screen physically admissible escape generators through their residual block couplings and adjoint modules, infer the completion algebra, and only then validate the selected circuit at finite depth. Residual algebra therefore provides not only an explanation for why a SP ansatz can fail, but also a constructive route for determining how it should be modified.

\bibliographystyle{apsrev4-2}
\bibliography{references_final}

\end{document}